\documentclass[twocolumn]{aastex63}
\usepackage{rotating}
\usepackage{graphicx}

\received{July 20, 2020}

\shorttitle{ZBLLAC - A spectroscopic database of BL Lacertae objects.}
\shortauthors{Landoni et. al.}

\graphicspath{{./}{figures/}}

\begin{document}

\title{ZBLLAC: A spectroscopic database of BL Lacertae objects.}

\correspondingauthor{Marco Landoni}
\email{marco.landoni@inaf.it}

\author[0000-0001-5570-5081]{Marco Landoni}
\affiliation{INAF - Istituto Nazionale di Astrofisica. Osservatorio Astronomico di Cagliari. Via della Scienza 5, 09047 Selargius (CA) - ITA}
\affiliation{INAF - Istituto Nazionale di Astrofisica. Osservatorio Astronomico di Brera Via E. Bianchi 46 - 23807 Merate (LC), ITA}

\author[0000-0003-4137-6541]{R. Falomo}
\affiliation{INAF - Istituto Nazionale di Astrofisica. Osservatorio Astronomico di Padova, Vicolo Osservatorio 5 - 35122 Padova, ITA}

\author[0000-0002-2239-3373]{S. Paiano}
\affiliation{INAF - Osservatorio Astronomico di Roma, via Frascati 33, I-00040, Monteporzio Catone, Italy}
\affiliation{INAF - Istituto Nazionale di Astrofisica. IASF Istituto di Astrofisica e Fisica Cosmica, Via Alfonso Corti 12 - 20133 Milano, ITA }

\author{A. Treves}
\affiliation{Università degli Studi dell' Insubria
Via Valleggio 11, 22100 Como, ITA}
\affiliation{INAF - Istituto Nazionale di Astrofisica. Osservatorio Astronomico di Brera Via E. Bianchi 46 - 23807 Merate (LC), ITA}

\begin{abstract}

This paper describes the database of optical spectra of BL Lacertae (BLL) objects ZBLLAC available at the URL \texttt{https://web.oapd.inaf.it/zbllac/}.
At the present date, it contains calibrated  spectra for 295 BLL. For about 35\% of them, we report a firm measure of redshift $z$ while for 35 sources we set a lower limit on $z$ based on the detection of intervening absorption systems, mainly ascribed to Mg II ($\lambda$2800 $\textrm{\AA})$. 
We report here on the architecture of the database and on its web front-end that permits to filter, query and interactively explore the data. 
We discuss some properties of the objects in the present dataset by giving the distribution of the redshifts and reporting on the detected emission lines, which turn out to be mainly forbidden and ascribed to [O II] ($\lambda$3737 $\textrm{\AA})$ and [O III] ($\lambda$5007 $\textrm{\AA})$. Finally, we discuss on intervening absorption systems detected in 35 BLLs that allow us to set lower limits to their distance.

\end{abstract}

\section{Introduction} \label{sec:intro}

BL Lacertae (BLL) are a peculiar class of low power ($L  \sim 10^{43}$ erg sec$^{-1}$) Active Galactic Nuclei (AGN) whose relativistic jet, generated by the accretion of matter onto a supermassive black hole, is closely aligned with the observer's line of sight \citep[e.g.][]{br78}. In this condition, the radiation produced by the jet is boosted due to the relativistic Doppler aberration, and in most cases dominates the Spectral Energy Distribution (SED) at almost any wavelength of the electromagnetic spectrum (see e.g. \cite{maraschi, ghisellini, urry}). A peculiar spectroscopic characteristic of BLL is that, at variance with other active galactic nuclei, the emission lines are absent or extremely weak (see e.g. \cite{falomorev} for a recent review) and the boosted non-thermal continuum, in most of the cases, outshines the contribution from the starlight of the host galaxy, which is typically a giant Elliptical with $M_v$ $\sim$ -22.50 \citep{sbarufattihst}. More generally, BLL are a subclass of a larger parent population called Blazars, which encompass also more powerful sources (namely Flat Spectrum Radio Quasars, FSRQ) with bolometric luminosity of the order of $10^{47}$, $10^{48}$ erg sec$^{-1}$ and optical spectra characterized by broad emission lines typical of quasars, suggesting the presence of a radiatively efficient accretion disk \citep{sha}.

The quasi-featureless continuum exhibited by BL Lacs is a characteristic that made them rather elusive since the determination of the redshift, which is a fundamental parameter to determine the distance and derive physical quantities, is in many cases hindered.  Historically, various groups made huge efforts to secure spectroscopic data with small-medium sized telescopes aiming to increase the number of BLL for which a firm determination of z was assessed, although for many objects this was limited to bright or moderately beamed sources due to the modest signal-to-noise ratio reachable with the available instrumentation (e.g. \cite{miller1,sti88, sti89,falomo1, falkot}).

Many surveys from radio band (e.g. 1 Jy radio catalog, \cite{stickel91}) to X-Ray missions (for instance the Einstein Observatory, ROSAT) allowed to increase the sample of known candidate BLLs stimulating spectroscopical follow ups in the optical band, that were carried out with various optical telescope facilites (see e.g. \cite{stocke85, lawrence96} and reference therein). The availability of deep optical surveys like the Sloan Digital Sky Survey (SDSS) also contributed to enlarge the number of BLLs. In particular, \cite{plotkin} by combining data from SDSS and radio catalogs discovered about 700 BLLs candidates and constrained the $z$ for 30\% of them.
More recently at high energies, the advent of FERMI mission \citep{fermi} showed that BLLs represent the dominant extragalactic population in gamma-rays and, intriguingly, that for most of them only poor quality spectroscopical data were available. Various methodologies, based on multi wavelength data (e.g. \cite{massarocolor, massarougs1,massarougs2,nori, massarosdss,dabruscowise, massaro15,massaro16,massarowise,pai17asso} and reference therein), developed to associate low-energy counterparts of objects detected by FERMI further encouraged spectroscopic campaigns (e.g. \cite{paggi,land15,massaroaa,ricci,crespo,crespoal,alvarezaj,paiano17gamma,marchesi,pena}). In this context, hundreds of BLLs spectra have been successfully secured (see e.g. the review from \cite{massaro16}) but only for the brightest targets the redshift was measured. In fact, for the faintest sources or extremely beamed BLLs, only observations with large aperture optical telescope (like the Very Large Telescope, Keck Telescope and Gran Telescopio CANARIAS) equipped with state-of-the-art instrumentation could allow to secure high signal-to-noise ratio spectra of these targets and reveal weak spectral featuress (intrinsic or intervening) to firmly determine their redshift (see e.g. \cite{sbarufattiaj, sba06,sbarufattilast, land13,sandrinelli,land15,landmas,land18,shaw,pita,paiano17gamma,paiano17tev,paianoneutrino,paiano19}).

The results of all these spectroscopic campaigns are frequently summarised across multi-wavelength catalogues in the literature (e.g. Roma BZCat, TevCAT see \cite{tevcat, bzcat}) but none of them allows scientists to access the fully calibrated spectra of the sources with a homogeneous set of figures that permit to identify the spectral features detected in the spectra and then reuse the data for many different scientific aims. Motivated by these facts, we developed a new web-based database of BL Lac objects, namely ZBLLAC (\texttt{https://web.oapd.inaf.it/zbllac/}), able to act as an online hub where optical spectra secured in the context of different publications with heterogeneous instrumentation are stored and made available to the community. 

We present in this paper the properties of the database and of the web application. ZBLLAC includes a smart representation of the data to provide a facilitated access to the basic information of each source and, specifically, to the machine readable 1-D calibrated spectra along with a PDF figure that shows the spectroscopic identification of firmly detected emission or absorption lines. The paper is organized as follows: in Section 2 we give an overview of our database and the website while in Section 3 we detail on the format of our data. In Section 4 we discuss properties of the dataset and we report our conclusions and future perspectives in Section 5.

\section{The ZBLLAC spectroscopic database}
At the time of writing of this work, the ZBLLAC database contains 337 objects considered as BLL or BLL candidates in the literature. The sources were selected among heterogeneous criteria such as their detection in $\gamma$-rays at GeV or TeV band (see e.g. \cite{fermi,paiano17gamma}, WISE infrared colour (\cite{massaro12, dabrusco19, demenezescat, menezes}) or the properties of their broadband SED (\cite{padovani95a, padovani95b, costamante02, costamante20}). According to the properties of their  optical spectra in ZBLLAC, we labeled 295 objects as BLL by adopting spectral criteria based on the absence of emission features or, if detected, on the value of the EW, luminosities and broadness of the lines. The remaining 42 targets exhibit spectra dominated by broad and intense emission lines, suggesting to us an alternative classification and for this reason they have not been considered as BLL.

For each object we give $\alpha$, $\delta$, catalog name, redshift, magnitude and  provide a flux calibrated spectrum\footnote{Data were made available to us in electronic form directly from the Authors. For further details see \texttt{https://web.oapd.inaf.it/zbllac/refindex.html}}, dereddened for Galactic extinction. The spectra are available both  in text format and with a PDF figure which reports both the flux calibrated and normalized spectra. The main detected features, if present, are  marked and identified (see examples of Figure 1). We also note that for 37 BLLs more than one spectrum, secured in different epochs and with different instrumentation, is reported in the database.

\begin{figure*}[]
\includegraphics[width=18cm, angle=-90]{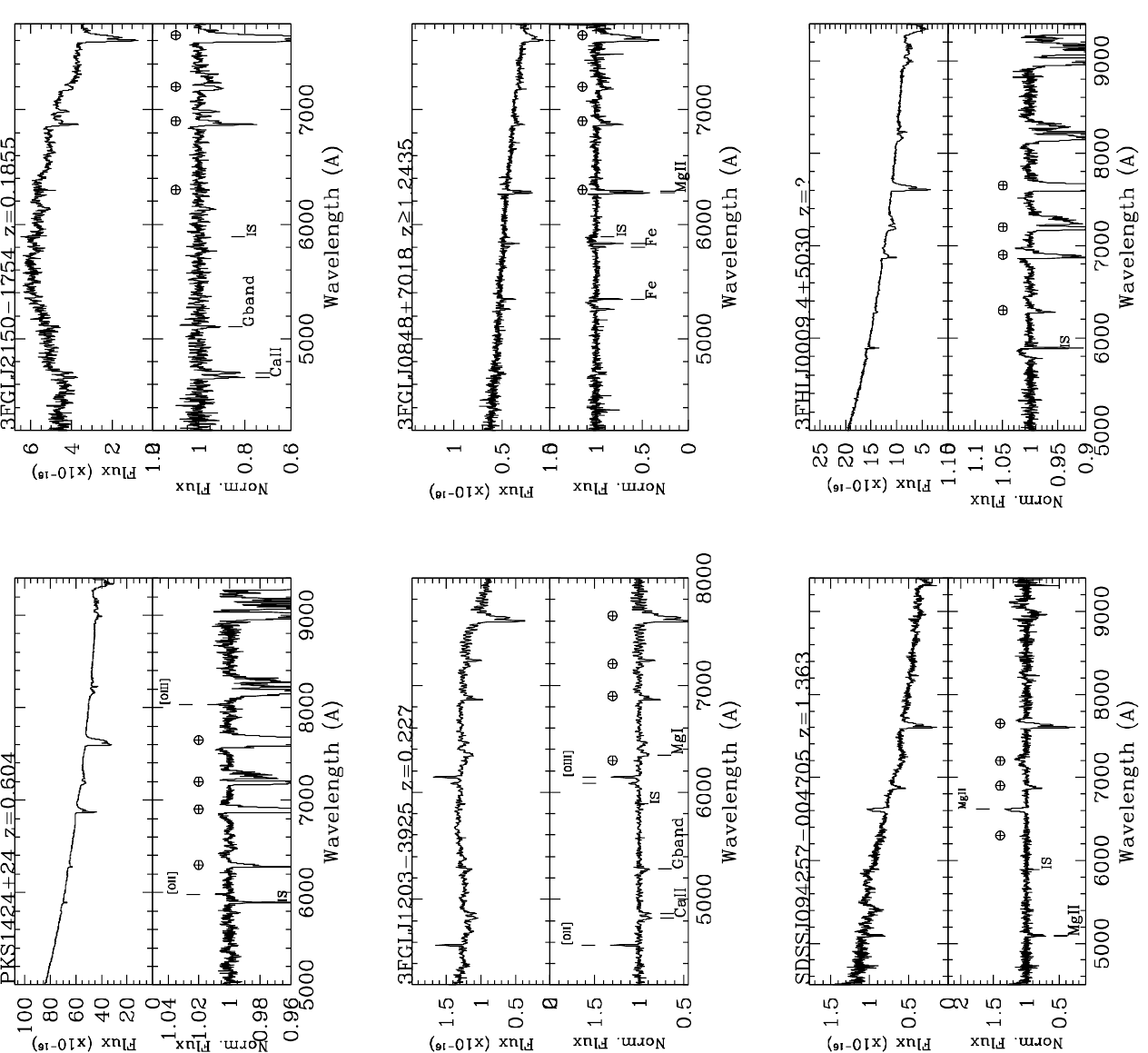}
\caption{Examples of spectra contained in the ZBLLAC database.
PKS1424$+$24: redshift derived from weak forbidden lines ascribed to [O II] and [O III] \citep{paiano17tev}.
3FGL J2150$-$1754: The $z$ follows from absorption lines ascribed to the host galaxy \citep{paiano17gamma}.
3FGL J1203$-$3925: The redshift derives from emission lines and features from the host galaxy \citep{herazo, marchesini}.
3FGL J0848$+$7018: The only apparent lines are due to intervening Mg II and Fe absorption systems, which yields a lower limit to the redshift \citep{paiano17gamma}.
SDSS J0942257$-$004705: The redshift is constrained through emission line from Mg II and absorption line from intervening system \citep{land18}.
3FHL J0009.4$+$5030: The continuum is completely featureless and the redshift is unknown \citep{paiano20}.
}
\end{figure*}

The web interface of ZBLLAC is reported in Figure 2. The user can retrieve and interactively explore the data through the Spectroscopic Database page, that shows by default the full list of sources present in ZBLLAC. For each of them, the application displays basic information and a set of buttons to download the spectrum and the annotated PDF figure. When more then one observation is associated to the same object, a further button  is displayed allowing to select the spectrum, or figure, to download among all the available ones for the source. Finally, we implemented a Search Panel, shown on top of the page as reported in Figure 2, that permits to actively filter the database according to various criteria such as coordinated within a radius, target name or redshift range.

\section{Data format and technological aspects} \label{sec:ict}

\begin{center}
\begin{figure*}
\hspace{-0.6cm}
\includegraphics[width=19cm]{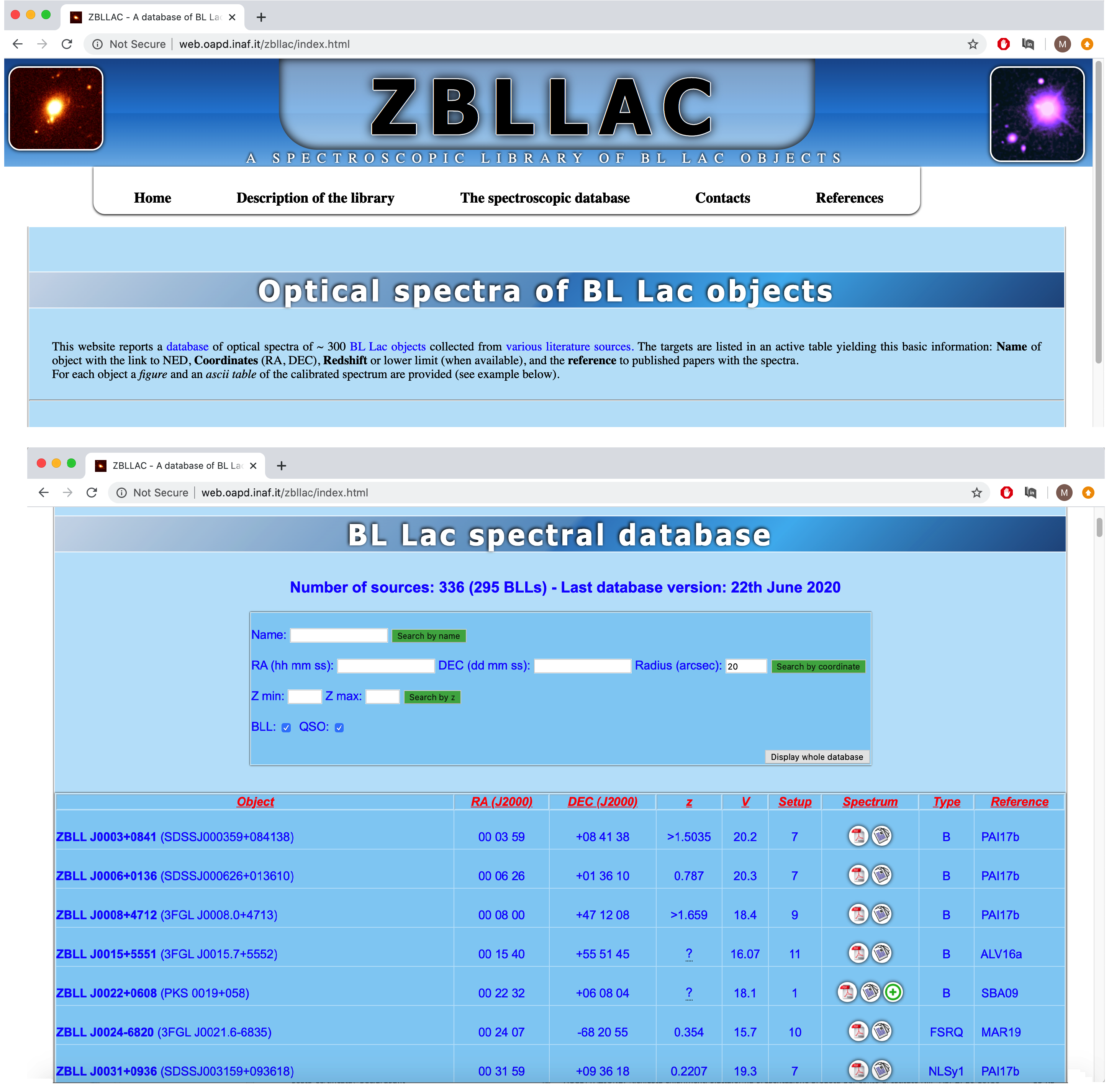}
\caption{The ZBLLAC database web interface accessible at the link \texttt{https://web.oapd.inaf.it/zbllac/}. Upper panel: Home page of the website. Lower panel: A small portion of the spectral database page that shows the sources present in ZBLLAC and the Search Panel to select the  objects on the basis of coordinates, redshift or literature name.}
\end{figure*}
\end{center}

\begin{figure*}[]
\includegraphics[width=17.0cm]{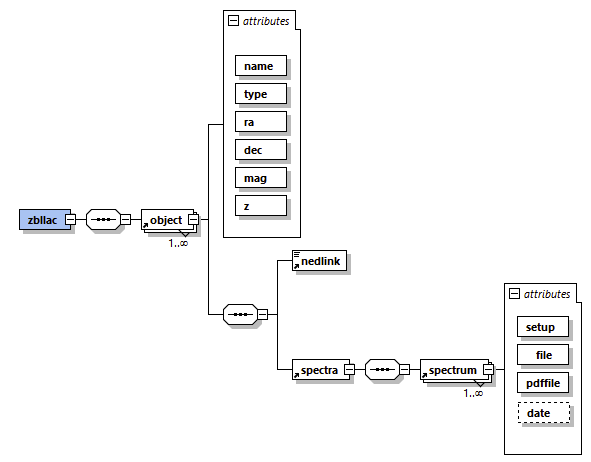}
\centering
\caption{Schematic of the XML-based database adopted for the ZBLLAC database.}
\end{figure*}

To store the data, we used the eXtensible Markup Language (XML\footnote{http://www.w3.org.}) standard that allows to define a proper data structure and model combining the possibility to easily query the database while being both machine and human readable. More specifically, the XML is a \textit{markup language} that defines a set of rules for encoding plain-text documents in which the data are marked by tags or attributes (see \cite{bosak} for a full review). 
Data within XML documents are organized using a tree-like data structure, where each node may posses one or more leafs.

We decided to adopt a representation of the data by using the XML scheme reported in Figure 3. The root node of the structure is \texttt{zbllac} which contains, as leafs, the set of all object of ZBLLAC database\footnote{The up-to-date version database can be download at this link : \texttt{https://web.oapd.inaf.it/zbllac/zbllac.xml}}. Each $\texttt{object}$ (see Figure 3) contains as attributes all the relevant information to identify the source (name, coordinates, etc.) and two mores nodes: the first one, named \texttt{nedlink}, contains the link to the object's NED page while the second (\texttt{spectra}) harbors a set of $\texttt{spectrum}$ nodes where the information about each observation and the relative HTTP links to download the data are saved. To further illustrate the organisation of the data, we show in Figure 4 a sample node of our XML file for the object $\texttt{PKS 1553+113}$ for which three different spectra have been secured.

For what concerns the website backend,  we made use of standard PHP pages coupled with XQuery and XPath\footnote{\texttt{ibm.com/developerworks/library/x-xpathphp/index.html}} protocols to retrieve the data throughout the XML file. 
\begin{figure*}[]
\includegraphics[width=18.9cm]{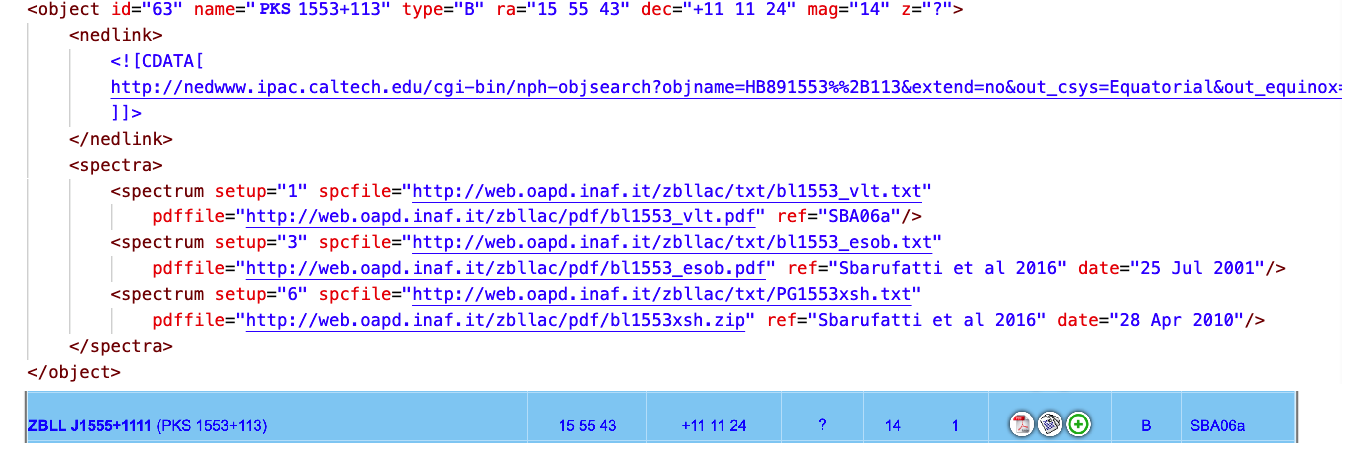}
\centering
\caption{An example of the XML representation of the data for the object PKS 1553+113. For this source, three different spectra have been secured. The attributes $\texttt{pdffile}$ and $\texttt{spcfile}$ contain the full link on our website to download the figure and the 1-D calibrated spectra of the source.
The lower blue panel shows how the row related to the source would be displayed on the ZBLLAC website.}
\end{figure*}


\section{Spectral properties of BLL} \label{sec:results}
\subsection{General properties}
The dataset of BLL, that currently encompass 295 targets, can be retrieved using the Spectroscopic database page (see Figure 2) by selecting the flag BLL.

For 103 ($\sim$ 35\% ) objects intrinsic spectral features are revelead, allowing to firmly measure the redshift. In details, for 31 objects only emission lines are detected in their spectra while in 55 BLLs only features from the host galaxy (mainly ascribed to Ca II and G Band) are present. In 17 cases, both emission and absorption features are revelead on the same spectrum.    \begin{figure}[]
\includegraphics[width=8cm]{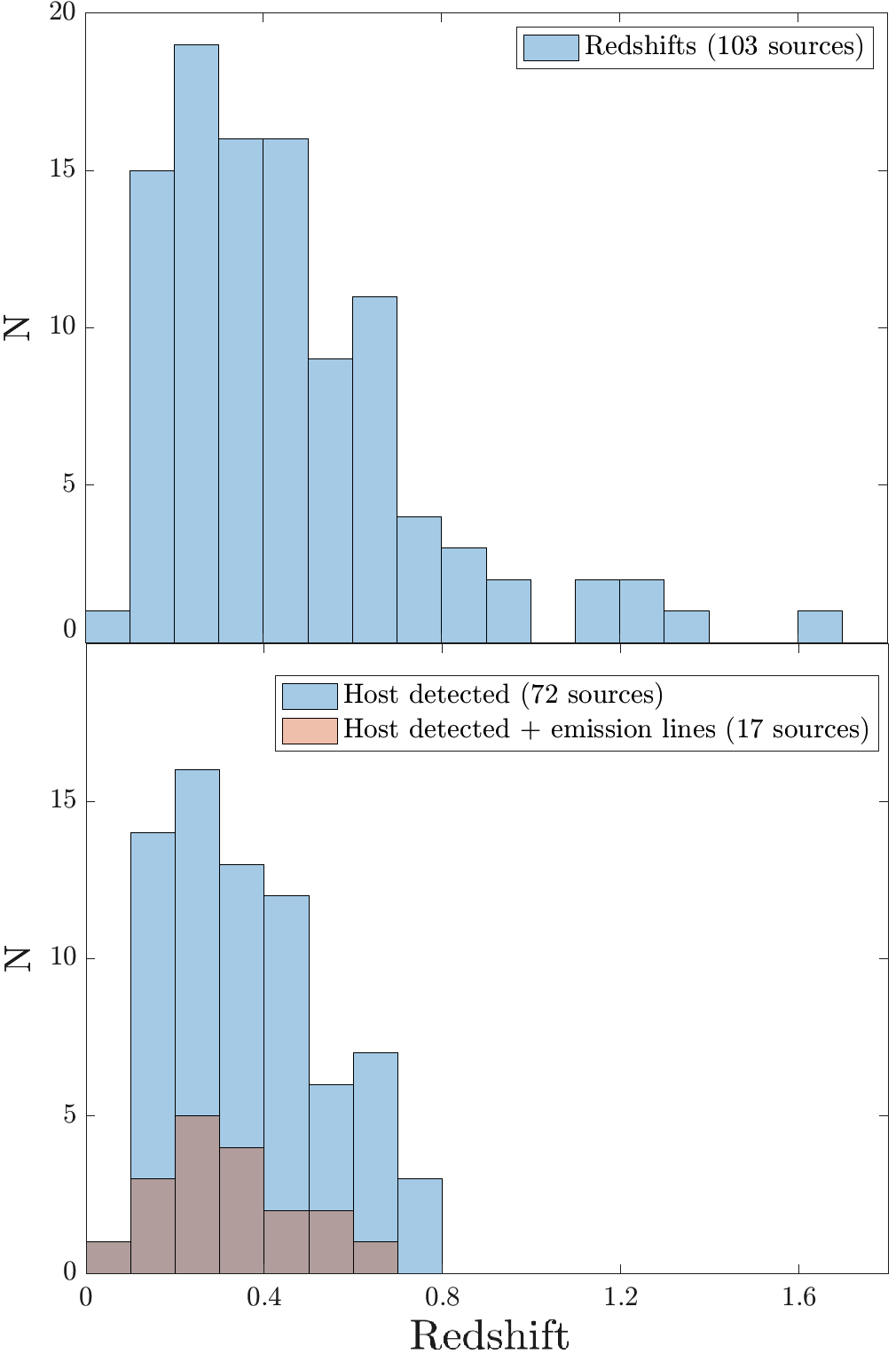}

\centering
\caption{Distribution of the redshift of 103 BLL objects  from the ZBLLAC database. Upper panel: distribution of $z$ of the whole dataset. Lower panel (pale blue): redshift of  the 72 sources in which the spectral features from the host galaxy have been detected. We report in shaded red the distribution of $z$ for 17 BLLs in which both emission and absorption lines have been revelead.}
\end{figure}
 The median value of $z$ is 0.40, ranging between 0.071 and 1.636. The distribution of the redshifts is given in Figure 5 where we also report an histogram of $z$ for objects in which the host galaxy has been detected. As shown in Figure 5, the determination of the redshift for object at $z \gtrsim 0.80$ is assesed only through emission lines since the features from host galaxy (e.g. Ca II $\lambda\lambda$ 3934-3968) start to move outside the covered spectral range (that for objects in ZBLLAC is, on average, between 4000$\textrm{\AA}$ and 8000$\textrm{\AA}$). 

Finally, for 35 BLLs that do not show intrinsic features we detected intervening absorption systems along the line of sight allowing to establish a firm lower limit to their redshift (see Section 4.3).

\subsection{Emission lines properties}
We detected emission features in 48 BLLs. The line identification and luminosities are given in Table 1. 

In 28 cases, the only observed lines are narrow forbidden transition ascribed to [O II] ($\lambda$ 3934 $\textrm{\AA}$) and [O III] ($\lambda$ 5007 $\textrm{\AA}$) while broad spectral features, mainly associated to Mg II ($\lambda$ 2800) and C III] ($\lambda$ 1908), are revealed in just 8 targets. We also note that for only 4 cases broad and narrow emission lines are present on the very same spectrum. 
These facts may suggest that, in the majority of the cases, there is no trace of the broad line region possibly meaning that either the physical conditions for its appearance are absent, or that the lines are so broad and swamped by the continuum that they are not detected. 
We report in Figure 6 the distribution of the luminosity of emission lines ascribed to [O II] ($\lambda$ 3934 $\textrm{\AA}$) and [O III] ($\lambda$ 5007 $\textrm{\AA}$). The median luminosity for [O II] is $1.7 \times 10^{41}$ erg s$^{-1}$ while in the case of [O III] we found $1.3 \times 10^{41}$ erg s$^{-1}$. 

Following the same approach described in \cite{paiano20}, we compared our luminosities of [O II] and [O III] with those measured on spectra of low redshift QSOs (\cite{shen}, with similar luminosity on the continuum assuming a beaming factor $\delta \sim$ 10) finding that their mean values are roughly similar. This result is in agreement with \cite{paiano20} but in this case our conclusions are based on a dataset which is significantly larger. 

\begin{figure}[]
\includegraphics[width=8cm]{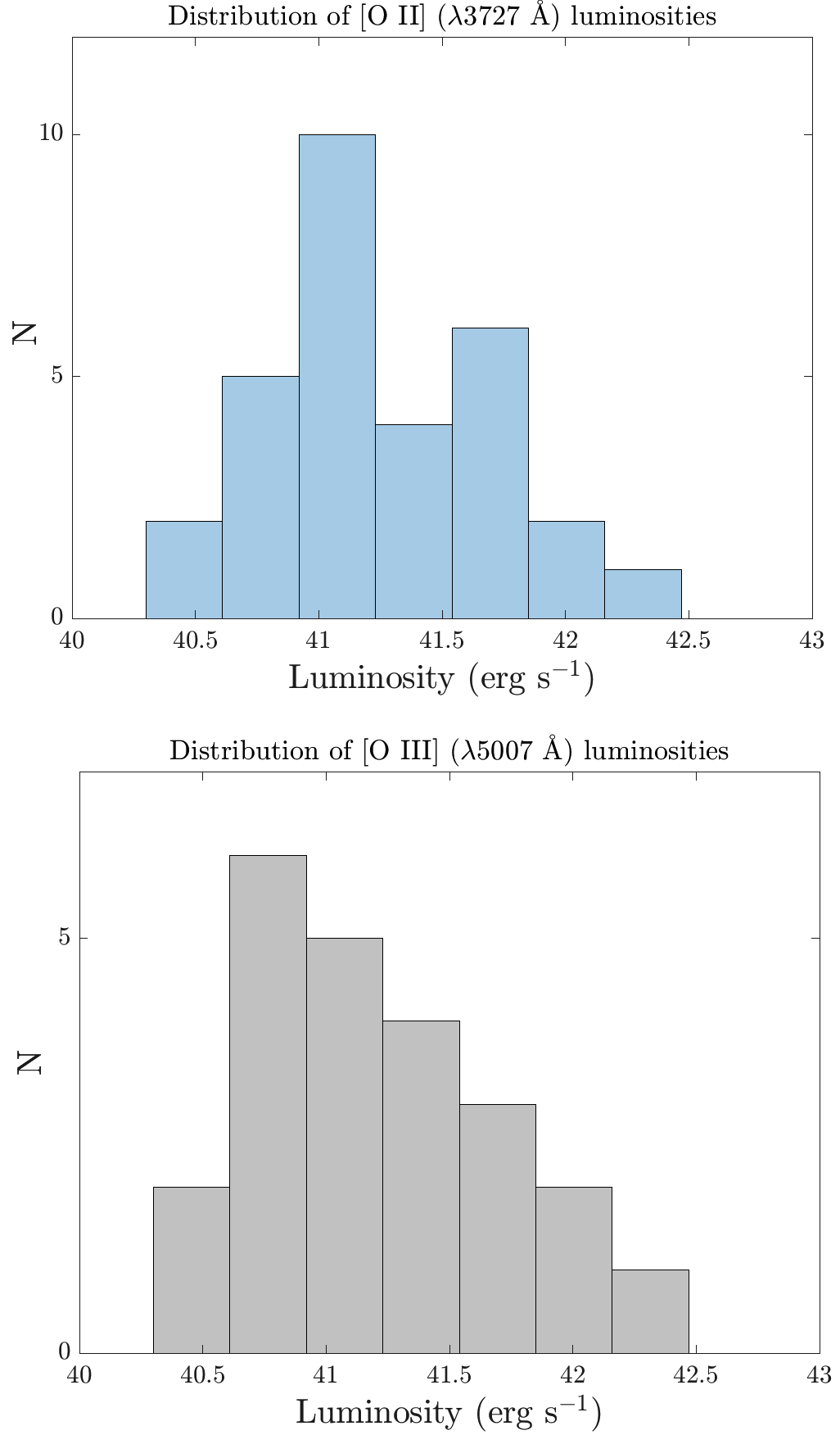}
\centering
\caption{Distribution of the  luminosity of emission lines from [O II] ($\lambda$ 3934 $\textrm{\AA}$) and [O III] ($\lambda$ 5007 $\textrm{\AA}$) detected in the spectra of BLLs in the ZBLLAC database.}
\end{figure}

\subsection{Intervening absorption systems}
In the spectra of BLLs, absorption lines could arise intrinsically from the host galaxy, yielding directly to the determination of $z$, or from intervening system if cool gases cloud structure is intercepted along the line of sight. In this case, the detection of an intervening absorption gives a robust lower limit to the redshift of the source. We detected those systems in 35 objects and we report our measurements on Table 2. 
In the wavelength range covered by our collection of spectra, the main absorption are those related to the Mg II doublet transition ($\lambda\lambda$ 2796-2803 $\textrm{\AA}$), when the redshift of the absorber is between $0.40 \leq z \leq 1.9$. In fact, in 30 cases we reveal spectral lines ascribed to Mg II ($\lambda$2800 $\textrm{\AA}$) that allow to set a lower limit to $z$. Furthermore, in three sources, both at redshift $z \gtrsim 2.00$, we detected the onset of Ly-$\alpha$ forest (see \cite{land18,paiano17gamma} for details) and further intervening system, at a lower redshift, associated to Mg II, C IV and Fe II (see Table 2). In a couple of targets, features arising from Ca II ($\lambda$ 3934$\textrm{\AA}$) are detected in intervening systems along the line of sight. 

The median value of our lower limits is $z \sim 0.64$ and we report in Figure  7 their distribution.
\begin{figure}[]
\includegraphics[width=8cm]{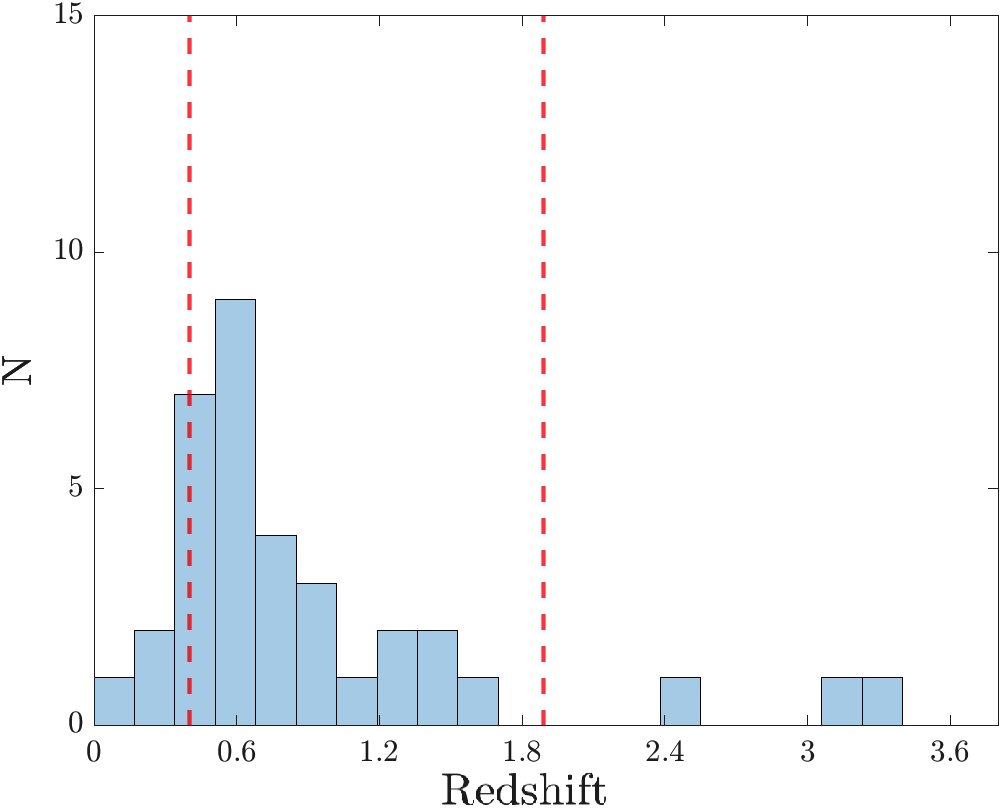}
\centering
\caption{Distribution of the redshift lower limits of 35 BLLs. The two red vertical bars show the redshift range $0.35\lesssim$ z $\lesssim 1.90$ in which our spectral coverage allow to detect intervening absorption lines from Mg II.}
\end{figure}
 The peak around $z \sim 0.60$ is related to our spectral range, where the probability of detection of Mg II is maximised. We note that lower limits for sources at $z \leq 0.40$ are in one case still ascribed to Mg II (ZBLL J0816$-$1311) because data has been obtained with ESO X-SHOOTER \citep{vernet} that provides increased spectral coverage in the UV \citep{pita}, while the other two cases are associated to intervening of Ca II ($\lambda 3934 \textrm{\AA}$). 

Finally, it is worthy to comment on the remaining 157 BLLs that still appear to be featureless. In our data, the total number of absorbers of Mg II, considering both multiple systems and  those found in BLLs with $z$, is 46. 
However, according to \cite{zhu}, \cite{land13} the expected number of detected Mg II systems in our dataset in the redshift range $0.40 \leq  z \leq 1.90$ (with EW $\geq 1.00 \textrm{\AA}$) should be of the order of $\sim $ 100. This consideration suggests that, in order to cope with this statistic, the 157 featureless BLLs should lie statistically at redshift $z \lesssim 0.70$ (see also \cite{paiano20} for similar conclusions).

\section{Concluding remarks}
We described the ZBLLAC database that currently contains optical spectra for 295 BLLs. We discussed the spectroscopic properties of the objects in this dataset finding that for 35\% of them intrinsic spectral features are revealed, allowing to solidly measure the redshift. We reported on 35 targets in which, by detecting intervening absorption systems, we set tight lower limits on their $z$. On the basis of the absence of absorption lines ascribed to Mg II in 157 featureless object, we statistically suggest that they should lie at low redshift $z \lesssim 0.70$.

The ZBLLAC spectroscopic database of BL Lac objects is an ongoing and still growing project. We encourage other groups to contribute by sharing their own published data. Instructions for joining our project and contributing to  the dataset can be found at \texttt{https://web.oapd.inaf.it/zbllac/Instructions.pdf}.

\section{Acknowledgements}
We thank B. Sbarufatti (Penn State University, USA),  F. Massaro (University of Torino), N. Crespo (ESA), E. Marchesini (Universidad Nacional de La Plata, Buenos Aires), A. Paggi (University of Torino), H. Pena-Herazo (Instituto Nacional de Astrofísica Óptica y Electrónica, Mexico), F. Ricci (Pontificia Universidad Catolica de Chile) ,  and S. Pita (Universite Paris Diderot) for their help in enriching the ZBLLAC database by sharing data.    
\newpage

\clearpage

\begin{longtable*} {|l|l|l|l|l|l|c|}

\hline
\textbf{Source}  & \textbf{$z$} & \textbf{Line}   & \textbf{EW $\AA$} & \textbf{FWHM (km/s)}  & \textbf{$L$ (erg/s)} & \textbf{Type}  \\
\hline

ZBLL J0035+5950       &0.467   &[O II] ($\lambda$3737)   &0.3  &350 &$9.7\cdot10^{40}$ & N  \\
\hline

ZBLL J0050-0929         &0.635   &[O II] ($\lambda$3727) &0.5  &600 &$1.0\cdot10^{41}$ & N \\
            &        &[O III] ($\lambda$5007) &0.6  &450  &$9.4\cdot10^{40}$ & N \\
            &        &H$\alpha (\lambda$6563) &1.6  &1300  &$1.6\cdot10^{41}$  & N\\
\hline
ZBLL J0048+4223  &0.302  &[O II] ($\lambda$3727)  &1.4  &920 &$4.6\cdot10^{40}$  & N\\
            &        &[O III] ($\lambda$5007) &1.2  &350 &$4.2\cdot10^{40}$ &N  \\
\hline

ZBLL J0158+0101  &0.4537 &[O III] ($\lambda$5007) &5.0  &350 &$7.0\cdot10^{40}$ & N \\
\hline

ZBLL J0303-2407      &0.2657 &[O III] ($\lambda$5007) &0.2  &300 &$8.5\cdot10^{40}$ & N \\
           &       &H$\alpha (\lambda$6563)  &0.2  &300 &$7.7\cdot10^{40}$ & N \\
            &       &[N II] ($\lambda$6585)  &0.2  &500 &$1.0\cdot10^{41}$ &N \\
           &       &[O II] ($\lambda$3727)        &0.1  &450 &$1.2\cdot10^{41}$ &N \\

    \hline
ZBLL J0301-1652  &0.278 &[O III]$ (\lambda$5007) &1.5  &400 &$3.4\cdot10^{40}$  &N\\
    \hline

\hline
ZBLL J0305-1608  &0.312  &[O II] ($\lambda$3727)   &2.0  & 800 &$1.7\cdot10^{41}$ &N \\
\hline

\hline
ZBLL J0316-2607  &0.443 &[O II] ($\lambda$3727) &0.52  &1600 &$1.3\cdot10^{41}$ &N \\
    \hline

ZBLL J0340-2119        &0.223   &[O II] ($\lambda$3727) &1.5  &1500 &$2.9\cdot10^{40}$ &N \\
            &        &[O III] ($\lambda$4960) &0.6  &1200 &$2.6\cdot10^{40}$  &N\\
\hline
ZBLL J0428-3756        &1.105   &C III] ($\lambda$1908)  &2.3  &3500 &$6.2\cdot10^{42}$ &B \\
            &        &Mg II ($\lambda$2800)   &3.0  &4500 &$6.3\cdot10^{42}$ &B \\
            &        &[O II] ($\lambda$3727)  &0.5  &650  &$9.2\cdot10^{41}$  &N \\
\hline
ZBLL J0509+0542       &0.3365 &[O II] ($\lambda$3737)   &0.07  &500 &$1.0\cdot10^{41}$ &N \\
                 &       &[O III] ($\lambda$5007)  &0.05  &600 &$9.2\cdot10^{40}$ &N \\
                 &       &[N II] ($\lambda$6585)   &0.05  &300 &$6.7\cdot10^{40}$ & N\\
    \hline

ZBLL J0550-3216         &0.068   &[N II] ($\lambda$6585)   &0.8  &550   &$2.0\cdot10^{42}$ &N\\
\hline

ZBLL J0757+0956         &0.266  &[O II] ($\lambda$3727) &0.6  &850 &$1.4\cdot10^{41}$  &N\\
                 &       &[O III] ($\lambda$5007) &0.9  &1100  &$1.9\cdot10^{41}$ &N \\

\hline
ZBLL J0811+0146          &1.148  &C III] ($\lambda$1908)   &1.0   &3000 &$2.0\cdot10^{42}$ &B \\
             &       &Mg II ($\lambda$2800)   &1.5   &4000 &$3.0\cdot10^{42}$  &B\\
    \hline

ZBLL J0820-1259               &0.539  &[O II] ($\lambda$3727)  &1.2   &1000 &$5.8\cdot10^{41}$ &N \\
             &       &H$\beta (\lambda$4862) &0.5  &850 &$1.9\cdot10^{41}$ &N \\
             &       &[O III] ($\lambda$5007)  &2.5  &650 &$8.4\cdot10^{41}$ &N \\
\hline

%

ZBLL J0930+5132 &0.1893  &[O III] ($\lambda$4960)  &1.0 &450 &$1.0\cdot10^{41}$ &N \\
\hline

ZBLL J0942-0047  &1.363 &C III] ($\lambda$1908) &4.8 & 200 &$4.5\cdot10^{41}$ &N \\
            &         &Mg II ($\lambda$2800)    &5.0  &1600 &$1.0\cdot10^{43}$  &N\\
    \hline
ZBLL J1008-3139    &0.534   &[O II] ($\lambda$3727)   &1.0  & 600 &$2.3\cdot10^{41}$  &N\\
\hline

ZBLL J1012+0631     &0.727  &Mg II$ (\lambda$2800)  &0.5     &1200  &$4.0\cdot10^{41}$ &N \\
            &       &[O II] ($\lambda$3727) &0.3     &900  &$5.0\cdot10^{41}$ &N \\
    \hline
    ZBLL J1046+5449  &0.252 &[O III] ($\lambda$5007) &4.0  &700 &$1.2\cdot10^{41}$ &N \\
    \hline

    ZBLL J1049+1548  &0.326 &[O II] ($\lambda$3727) &0.3  &1200 &$1.2\cdot10^{41}$ &N \\
    \hline

    ZBLL J1058-8003 &0.581 &Mg II ($\lambda$2800) &1.2  &2500 &$4.6\cdot10^{42}$&B  \\
        &      &[O III] ($\lambda$4960) &0.5  &600 &$8.3\cdot10^{41}$&N  \\
        &      &[O III] ($\lambda$5007) &1.4  &600 &$2.5\cdot10^{42}$&N  \\
\hline
    ZBLL J1117+2014  &0.140 &[O II] ($\lambda$3727) &0.8  &1200 &$5.2\cdot10^{40}$ &N \\
    \hline

    ZBLL J1203-3926  &0.227 &[O II] ($\lambda$3727) &2.4  &600 &$6.1\cdot10^{40}$ &N \\
         &   &[O III] ($\lambda$4960) &1.4  &800 &$3.7\cdot10^{40}$&N  \\
         &   &[O III] ($\lambda$5007) &2.82  &550 &$7.3\cdot10^{40}$&N  \\

\hline
   
    ZBLL J1215+0732  &0.137 &H$\alpha (\lambda$6563) &1.3  &600 &$3.3\cdot10^{40}$ &N \\
    \hline

    ZBLL J1217+3007 &0.129 &[O II] ($\lambda$3727) &0.2 &600 &$6.5\cdot10^{40}$ &N \\
         &   &[O III] ($\lambda$5007) &0.2  &650 &$5.1\cdot10^{40}$&N  \\
   
    \hline
    ZBLL J1221+2813 &0.102 &[O III] ($\lambda$5007) &0.8  &600 &$4.8\cdot10^{40}$&N  \\
    \hline

    \hline
    ZBLL J1231+3711  &0.219 &[O II] ($\lambda$3727) &5.0  &1000 &$1.4\cdot10^{41}$&N  \\
    \hline
   
    ZBLL J1240+3445  &1.636 &Mg II ($\lambda$2800) &4.5  &3500 &$4.2\cdot10^{42}$&B  \\
    \hline

    ZBLL J1247+4423  &0.569 &[O III] ($\lambda$5007) &1.4  &400 &$1.8\cdot10^{41}$&N  \\
    \hline

    ZBLL J1259-2310  &0.481 &[O II] ($\lambda$3727) &0.9  &1100 &$4.5\cdot10^{41}$ &N \\
           &      &[O III] ($\lambda$5007) &0.4  &600 &$1.7\cdot10^{41}$ &N \\
    \hline

    ZBLL J1309+4305 &0.693 &[O II] ($\lambda$3727) &1.2  &800 &$2.0\cdot10^{42}$&N  \\
           &      &[O III] ($\lambda$5007) &0.5  &600 &$5.5\cdot10^{41}$ &N \\

\hline
    ZBLL J1427+2348  &0.604 &[O II] ($\lambda$3727) &0.1  &300 &$5.0\cdot10^{41}$&N  \\
          &      &[O III] ($\lambda$5007) &0.2  &400 &$1.1\cdot10^{42}$ &N \\
 \hline
ZBLL J1522-2730         &1.297   &Mg II ($\lambda$2800) &0.4  &2000 &$3.0\cdot10^{42}$ &B \\
\hline

ZBLL J1541+1414  &0.223  &[O III] ($\lambda$5007) &1.0  &500 &$3.4\cdot10^{40}$ &N \\
\hline

ZBLL J1626-7638  &0.1050 &[O I] ($\lambda$6302) &1.5  &1000 &$3.7\cdot10^{40}$&N  \\
           &       &[S II] ($\lambda6718-6732$) &2.0  &700 &$4.6\cdot10^{40}$ &N \\
\hline
ZBLL J1637+1314  &0.655 &[O II] ($\lambda$3727) &0.4  &800 &$1.5\cdot10^{41}$&N  \\
\hline

ZBLL J1704+1234  &0.452 &[O II] ($\lambda$3727) &2.3  &900 &$4.0\cdot10^{41}$&N \\
           &   &[O III] ($\lambda$4960) &0.7  &600 &$1.1\cdot10^{41}$&N  \\
           &   &[O III] ($\lambda$5007) &3.0  &600 &$5.0\cdot10^{41}$&N \\
\hline
ZBLL J1917-1921     &0.137   &[O II] ($\lambda$3727) &0.2  &600 &$3.5\cdot10^{40}$ &N \\
      &        &[O III] ($\lambda$4960) &0.2  &1000 &$2.4\cdot10^{40}$ &N \\
      &        &[O III] ($\lambda$5007) &0.5  &900 &$5.0\cdot10^{40}$&N  \\
\hline

\hline
ZBLL J2009-4849   &0.071   &H$\alpha (\lambda$6563) &0.3 &400 &$2.0\cdot10^{40}$ &N \\
\hline

ZBLL J2134-0153   &1.284   &C III] ($\lambda$1908) &1.2  &1800 &$2.2\cdot10^{42}$ &B \\
       &        &C II] ($\lambda$2326) &0.5  &1800 &$9.0\cdot10^{41}$&B  \\
      &        &Mg II ($\lambda$2800) &2.3  &3500 &$4.0\cdot10^{42}$&B \\
\hline

ZBLL J2152+1734       &0.870   &Mg II ($\lambda$2800) &2.7  &3000 &$1.2\cdot10^{42}$&B  \\
      &        &[O II] ($\lambda$3727) &2.2  &1100 &$9.5\cdot10^{41}$&N  \\

\hline
ZBLL J2209-0451  &0.3967 &[O II] ($\lambda$3727) &0.5  &300 &$7.7\cdot10^{40}$&N  \\
\hline

ZBLL J2225-1113   &0.997   &[O II] ($\lambda$3727) &1.8  &800 &$2.0\cdot10^{41}$ &N \\
\hline

ZBLL J2246+1544  &0.5965 &[O II] ($\lambda$3727) &0.9  &850 &$2.6\cdot10^{41}$&N \\
\hline

   ZBLL J2250+1749  &0.3437 &[Ne V] ($\lambda$3426) &1.5  &350 &$4.3\cdot10^{40}$ &N \\
         &    &[O II] ($\lambda$3727) &3.0  &500 &$9.5\cdot10^{40}$ &N \\
         &    &[O III] ($\lambda$4960) &1.0  &350 &$5.4\cdot10^{40}$&N  \\
        &    &[O III] ($\lambda$5007) &4.5  &500 &$2.3\cdot10^{41}$&N  \\

    \hline
    ZBLL J2349+0534  &0.419 &Mg II ($\lambda$2800) &4.6  &3000 &$6.6\cdot10^{41}$&B  \\
           &      &[O II] ($\lambda$3727) &3.0  &1000 &$3.5\cdot10^{41}$&N  \\
           &      &[O III] ($\lambda$4960) &1.6  &750 &$1.6\cdot10^{41}$ &N \\
           &      &[O III] ($\lambda$5007) &3.5  &700 &$3.5\cdot10^{41}$&N  \\
\hline
   ZBLL J2357-0152   &0.812 & Mg II ($\lambda$2800) &1.4  &1500 &$2.8\cdot10^{41}$ &N \\
    \hline
\caption{Properties of the emission lines detected in the spectra of BLLs that belong to the ZBLLAC database. }
    \end{longtable*}

\clearpage
\begin{longtable*} {|l|l|c|l|}

\hline
\textbf{Source}  & \textbf{Line}   & EW (\textrm{\AA})& \textbf{z$_{abs}$}   \\
\hline
ZBLL J0003$+$0841 & Mg II ($\lambda$ 2800) & 1.50 & 1.5035 \\
\hline
ZBLL J0008$+$4712 & Mg II ($\lambda$ 2800) & 2.00 & 1.659 \\
\hline
ZBLL J0033$-$1921 &  Mg II ($\lambda$ 2800) & 0.20 & 0.505 \\
\hline
ZBLL J0038$+$0012  & Mg II ($\lambda$ 2800) &0.70 & 0.80\\
\hline
ZBLL J0234$-$0628  &  Mg II ($\lambda$ 2800) & 7.00 & 0.63 \\
\hline
ZBLL J0251$-$1831  &  Mg II ($\lambda$ 2800) & 3.50 & 0.615 \\
\hline
ZBLL J0338$+$1302  &  Mg II ($\lambda$ 2800) & 3.00 & 0.382 \\
\hline
ZBLL J0441$-$2952 &  Mg II ($\lambda$ 2800) &  2.15 & 0.68 \\
\hline
ZBLL J0644$+$6038  & Mg II ($\lambda$ 2800) & 5.00 &0.581 \\
\hline
ZBLL J0649$-$3139 & Mg II ($\lambda$ 2800) & 3.00 & 0.563 \\
\hline
ZBLL J0816$-$1311 & Mg II ($\lambda$ 2800) & 0.15 & 0.2882 \\
&           Mg II ($\lambda$ 2800)       & 0.60 & 0.2336 \\
& Mg II ($\lambda$ 2800) & 1.00 & 0.1902\\

\hline
ZBLL J0848$+$7017 & Mg II ($\lambda$ 2800) & 11.30 & 1.2435 \\
\hline

ZBLL J1107$+$0222 & Mg II ($\lambda$ 2800) & 2.00 & 1.0735 \\
\hline
ZBLL J1129$+$3756  & Mg II ($\lambda$ 2800) & 9.10 & 1.211 \\
\hline

ZBLL J1231$+$0138 & Ly $\alpha$ (1216) & 15.00 & 3.140 \\
                  & Mg II ($\lambda$ 2800) & 6.00 & 2.004\\
                  & Fe II ($\lambda$ 2600) & 5.00 & 2.004\\
                & Mg II ($\lambda$ 2800) & 4.00 & 2.004\\
\hline
ZBLL J1223$+$0820 & Mg II ($\lambda$ 2800) & 0.90 & 0.7187 \\
\hline
ZBLL J1243$+$3627 & Mg II ($\lambda$ 2800) & 0.90 & 0.48 \\
\hline
ZBLL J1312$-$2350  & Mg II ($\lambda$ 2800) & 2.50 & 0.462 \\
\hline
ZBLL J1351$+$1114  & Mg II ($\lambda$ 2800) & 1.00 & 0.619 \\
\hline
ZBLL J1450$+$5201 & Ly $\alpha$ ($\lambda$ 1216) & 5.80 & 2.470 \\
                & C IV ($\lambda$ 1908) & 3.20 & 2.470 \\
                & C IV ($\lambda$ 1908) & 1.10 & 2.312\\
\hline
ZBLL J1511$-$0513 & Mg II ($\lambda$ 2800) & 2.10 & 0.451 \\
\hline
ZBLL J1540$+$8155 &Mg II ($\lambda$ 2800) & 0.60 & 0.672 \\
\hline
ZBLL J1730$-$0352  & Mg II ($\lambda$ 2800) & 7.50 & 0.776 \\
\hline
ZBLL J1955$-$1603  &  Mg II ($\lambda$ 2800) &3.00 & 0.638 \\
\hline
ZBLL J1959$-$4725  & Mg II ($\lambda$ 2800) & 2.30 & 0.519\\
\hline

ZBLL J1200+4009 & Ly $\alpha$ ($\lambda$ 1216) & 13.00 & 3.367 \\
 & Mg II ($\lambda$ 2800) & 2.50 & 1.484 \\
 & Fe II ($\lambda$ 2600) & 2.00 & 1.484 \\
 & Mg II ($\lambda$ 2800) & 4.50 & 1.142 \\
\hline
ZBLL J2107$-$4828 & Mg II ($\lambda$ 2800) & 4.30 & 0.519 \\
\hline
ZBLL J2115$+$1218 & Mg II ($\lambda$ 2800) & 4.00 & 0.497 \\
                 &Mg II ($\lambda$ 2800) & 0.90 & 0.525 \\
                 & Mg II ($\lambda$ 2800)& 0.90 & 0.633 \\
\hline
ZBLL J2139-4235 & Ca II ($\lambda\lambda$ 3934-3968) & 0.25 & 0.0087\\
\hline
ZBLL J2212$+$2759 & Mg II ($\lambda$ 2800) & 3.90 & 1.529\\
\hline
ZBLL J2236$-$1433  & Mg II ($\lambda$ 2800) & 0.70 & 0.490 \\
             & Mg II ($\lambda$ 2800) & 0.90 & 0.493 \\
\hline
ZBLL J2247$+$0000 & Mg II ($\lambda$ 2800) & 3.00 & 0.898 \\
\hline
ZBLL J2255$+$2410 & Mg II ($\lambda$ 2800) & 0.70 & 0.8633\\
\hline
ZBLL J2319$+$1612  & Mg II ($\lambda$ 2800) & 1.50 & 0.970 \\
\hline
ZBLL J2323$+$4210 & Ca II ($\lambda\lambda$ 3934-3968) & 0.50 & 0.267 \\
                  & Na I ($\lambda$ 5892) & 0.35 & 0.267\\

\hline
\caption{Properties of the intervening absorption lines detected in the spectra of BLL that allow to derive a redshift lower limit.}
    \end{longtable*}

\bibliography{zbllac}{}

\bibliographystyle{aasjournal}

\end{document}